\documentclass[aps,prd,showpacs,preprintnumbers]{revtex4}
\usepackage{amsmath,amssymb}
\usepackage{slashed}
\usepackage{graphicx}
\usepackage{subfigure}
\usepackage{color}
\newcommand\bef{\begin{figure}}
\newcommand\eef[1]{\label{fg:#1}\end{figure}}
\newcommand\beq{\begin{equation}}
\newcommand\eeq[1]{\label{#1}\end{equation}}
\newcommand\bea{\begin{eqnarray}}
\newcommand\eea{\end{eqnarray}}
\newcommand\bet{\begin{table}}
\newcommand\eet[1]{\label{tbl:#1}\end{table}}

\newcommand\fgn[1]{Figure \ref{fg:#1}}
\newcommand\eqn[1]{eq.\ (\ref{#1})}
\newcommand\scn[1]{Section \ref{sec.#1}}
\newcommand\apx[1]{Appendix \ref{sec.#1}}
\newcommand\tbn[1]{Table \ref{tbl:#1}}

\newcommand\jhep{{\sl J.\ H.\ E.\ P.\/}\ }
\newcommand\np{{\sl Nucl.\ Phys.\/}\ }

\newcommand\pr{{\sl Phys.\ Rev.\/}\ }

\newcommand\plt{{\sl Phys.\ Lett.\/}\ }

\DeclareMathOperator\arcsinh{arcsinh}

\newcommand\tr{\mathrm{Tr}\;}

\newcommand{\txt}{\textstyle}

\newcommand{\nt}{N_\tau}
\newcommand{\ns}{N_\sigma}
\newcommand{\ttc}{T/T_c}
\newcommand{\tc}{T_c}
\newcommand{\ti}{\tau_{\rm imp}}
\newcommand{\tmin}{\tau_{\rm imp}^{\rm min}}
\newcommand{\kt}{\kappa/T^3}
\newcommand{\ke}{\kappa_{\scriptscriptstyle E}}
\newcommand{\kb}{\kappa_{\scriptscriptstyle B}}
\newcommand{\ket}{\kappa_{\scriptscriptstyle E}/T^3}
\newcommand{\kbt}{\kappa_{\scriptscriptstyle B}/T^3}
\newcommand{\rom}{\rho_{\scriptscriptstyle T}(\omega)}

\newcommand{\etad}{\eta_{\scriptscriptstyle D}}
\newcommand{\vx}{\vec{x}}

\newcommand{\md}{m_{\scriptscriptstyle D}}
\newcommand{\mq}{m_{\scriptscriptstyle Q}}
\newcommand{\pq}{p_{\scriptscriptstyle Q}}
\newcommand{\mqi}{\frac{\textstyle 1}{\textstyle \mq}}
\newcommand{\mufit}{\mu_{\rm fit}}
\newcommand{\cf}{C_f}
\newcommand{\ds}{2 \pi \, T \, D_s}
\newcommand{\dsc}{D_s^c}
\newcommand{\dsb}{D_s^b}
\newcommand{\dscb}{2 \pi \, T \, D_s^{c,b}}
\newcommand{\zee}{Z_{\scriptscriptstyle EE}}
\newcommand{\raa}{R_{\scriptscriptstyle AA}}

\newcommand{\cosec}{{\rm cosec}}
\newcommand{\gbsq}{g_{\scriptscriptstyle B}^2}
\newcommand{\ltau}{L_\tau}

\newcommand{\gcont}{G^{\scriptscriptstyle \rm Cont}_{\scriptscriptstyle \rm norm}}
\newcommand{\glat}{G^{\scriptscriptstyle \rm Lat}_{\scriptscriptstyle \rm norm}}

\newcommand{\gelat}{G_{\scriptscriptstyle EE}^{\scriptscriptstyle \rm bare}}
\newcommand{\gern}{G_{\scriptscriptstyle EE}^{\scriptscriptstyle \rm renorm}}
\newcommand{\gfr}{G_{norm}}
\newcommand{\get}{G_{\scriptscriptstyle EE}}
\newcommand{\gbt}{G_{\scriptscriptstyle BB}(\tau)}
\newcommand{\ggfr}{\frac{\textstyle \get}{\textstyle \glat} (\tau)}

\newcommand{\puv}{\rho_{\scriptscriptstyle UV}(\omega)}
\newcommand{\pir}{\rho_{\scriptscriptstyle IR}(\omega)}
\def\dtk(#1){\int \frac{d^3 {#1}}{8 \pi^3}}
\def\om(#1){\mathcal{O}(\mq^{-#1})}

\begin{document}
\title{Temperature dependence of the static quark diffusion coefficient}
\author{Debasish Banerjee}
\email{debasish.banerjee@saha.ac.in}
\affiliation{Saha Institute of Nuclear Physics, HBNI, Kolkata 700064, India}
\affiliation{Homi Bhabha National Institute, Training School Complex, 
Anushaktinagar, Mumbai 400094, India}
\author{Saumen Datta}
\email{saumen@theory.tifr.res.in}
\affiliation{Tata Institute of Fundamental Research, Homi Bhabha Road,
  Mumbai 400005, India}
\author{Rajiv V. Gavai}
\email{gavai@tifr.res.in}
\affiliation{Indian Institute of Science Education and Research, 
  Bhauri, Bhopal 462066, India}
\author{Pushan Majumdar}
\thanks{Deceased.}
\affiliation{Indian Association for the Cultivation of Science,
  Raja S. C. Mullick Road, Kolkata 700032, India}

\begin{abstract}
  The energy loss pattern of a low momentum heavy quark in a
  deconfined quark-gluon plasma can be understood in terms of a
  Langevin description.  In thermal equilibrium, the motion can then
  be parametrized in terms of a single heavy quark momentum diffusion
  coefficient $\kappa$, which needs to be determined
  nonperturbatively. In this work, we study the temperature dependence
  of $\kappa$ for a static quark in a gluonic plasma, with a
  particular emphasis on the temperature range of interest for heavy
  ion collision experiments.
\end{abstract}
\pacs{12.38.Mh, 12.38.Gc, 25.75.Nq}
\preprint{TIFR/TH/22-36}

\maketitle

\section{Introduction}
\label{sec.intro}
The charm and the bottom quarks provide very important probes of the
medium created in the relativistic heavy ion collision
experiments. Since the masses of both of these quarks are much larger
than the temperatures attained in RHIC and in LHC, one expects these
quarks to be produced largely in the early pre-equilibrated state of
the collision. Heavy quark probes therefore provide a window to look
into the early stages of the fireball.

In particular, the nature of the interaction of the heavy quarks with
the thermalized medium is different from that of the light quarks. For
energetic jets, radiative energy loss via bremsstrahlung is expected
to be the dominant energy loss mechanism. For heavy quarks, the
radiative energy loss is suppressed in a cone of angle $\sim \mq/E$
\cite{dk}. For heavy quarks of moderate energy, $E \lesssim 2 \mq$,
collision with thermal quarks and gluons is expected to be the
dominant mechanism of energy loss \cite{mt,mustafa}.

Even if the kinetic energy of the heavy quark is $\mathcal{O}(T)$,
where $T$ is the temperature of the fireball, its momentum will be
much larger than the temperature. Its momentum is, therefore, changed
very little in a single collision, and successive collisions can be
treated as uncorrelated. Based on this picture, a Langevin description
of the motion of the heavy quark in the medium has been proposed
\cite{svetitsky}, \cite{mt,mustafa}. $v_2$, the elliptic flow
parameter, can then be calculated in terms of the diffusion
coefficient of the heavy quark in the medium. The diffusion
coefficient has been calculated in perturbation theory
\cite{svetitsky,mt}. While this formalism works quite well in
explaining the experimental data for $\raa$ and $v_2$ of the $D$
mesons (see \cite{annrev} for a review), the diffusion coefficient
required to explain the data is found to be at least an order of
magnitude lower than the leading order (LO) perturbation theory (PT)
result.

This is not a surprise {\em per se}, as the quark-gluon plasma is
known to be very nonperturbative at not-too-high temperatures, and
various transport coefficients have been estimated to have values very
different from LOPT. However, this makes it important to have a
nonperturbative estimate of the heavy quark diffusion coefficient. A
field theoretic definition of the heavy quark diffusion coefficient to
leading order of an $1/M$ expansion was given in \cite{ct, clm}.  The
next-to-leading order (NLO) calculation of the diffusion coefficient
in perturbation theory \cite{cm} was found to change the LO result by
nearly an order of magnitude at temperatures $\lesssim 2 \tc$. While
the NLO correction is in the direction suitable for explaining the
experimental data, the large change from LO to NLO indicates an
inadequacy of perturbation theory in obtaining a reliable estimate for
the diffusion coefficient in the temperature range of interest, and
makes a nonperturbative estimate essential.

The first nonperturbative results for $\kappa$, using the formalism of
\cite{clm} and numerical lattice QCD in the quenched approximation
(i.e., gluonic plasma), supported a value of $\kappa$ substantially
different from LOPT and in the correct ballpark for HIC phenomenology
\cite{prd11}. Of course, the plasma created in experiments is not a
gluonic plasma, and one needs a full QCD calculation for phenomenology;
but the fact that the quenched QCD result is of the right order of
magnitude gives strong support for the Langevin description of the 
heavy quark energy loss. Later works \cite{bielefeld, tum1, gflow}
improved on the calculations of \cite{prd11}.  In particular,
\cite{bielefeld,gflow} conducted a study at 1.5 $\tc$, and explored
various systematics in the numerical calculation of $\kappa$. The focus 
of Ref. \cite{tum1} was a comparison with perturbation theory, and
asymptotically high temperatures were explored.  Meanwhile, a
nonperturbative definition of the first correction to the static limit
was discussed in \cite{blaine}. Nonperturbative estimates of this
correction have recently been carried out \cite{bb,tum2}.

In this work we have carried out a study of the static quark momentum
diffusion coefficient $\kappa$ in the temperature range $\lesssim 3.5
\tc$, following the formalism of \cite{clm}. The focus here is on
studying the temperature dependence of $\kappa$ in the temperature
range of interest for the relativistic heavy ion collision
experiments. We extend the temperature range studied in \cite{prd11}
to cover the entire temperature range of interest to the heavy ion
community, and also improve the analysis technique, following
Refs. \cite{bielefeld} and \cite{bb}.  After explaining the formalism
and our calculational techniques in \scn{form} and \scn{tech},
respectively, we present the results of our study in \scn{results}.
Combined with the $1/M$ correction terms calculated in \cite{bb},
we can get the results for momentum diffusion coefficients for the
charm and the bottom in the plasma. We discuss these results in
\scn{summary}.

\section{Langevin formalism and nonperturbative definition of the
  momentum diffusion coefficient $\kappa$}
\label{sec.form}
In this section, we outline the formalism underlying our study. We first
define the Langevin formalism for the heavy quark energy loss, as
described in \cite{svetitsky}, \cite{mt, mustafa}, and then give a 
nonperturbative definition of the diffusion coefficient, following
\cite{ct,clm}.  

The heavy quark momentum is much larger than the system temperature T:
even for a near-thermalized heavy quark with kinetic energy $\sim T$,
its momentum $\pq \sim \sqrt{\mq \, T}$, where $\mq$ is the
heavy quark mass. Individual collisions
with the medium constituents with energy $\sim T$ do not change the
momentum of the heavy quark substantially if $\mq \gg T$.
Therefore, the motion of the heavy quark is similar to a Brownian
motion, and the force on it can be written as the sum of a drag term
and a ``white noise'', corresponding to uncorrelated random
collisions: \\ 
\beq 
\frac{d p_i}{dt} \ = \ - \etad \, p_i \ +
\ \xi_i(t), \qquad \langle \xi_i(t) \xi_j(t^\prime) \rangle \ = \ \kappa
\ \delta_{i j} \ \delta(t-t^\prime) .
\eeq{langevin} 
The momentum diffusion coefficient,
$\kappa$, can be obtained from the correlation of the force term: \\ 
\beq 
\kappa \ = \ \frac{1}{3} \ \int_{- \infty}^\infty dt \ \sum_i
\langle \xi_i(t) \, \xi_i(0) \rangle .
\eeq{force}
The drag coefficient, $\etad$, can be connected to the diffusion
coefficient using standard fluctuation-dissipation relations
\cite{kapusta}: \\ 
\beq 
\etad = \frac{\kappa}{2 \mq T}.
\eeq{fd}

In the leading order in an expansion in $\mqi$, the heavy quark
interacts only with the color electric field of the plasma. Therefore 
the momentum diffusion coefficient $\kappa$ can be obtained from
the electric field correlation function \cite{ct,clm}
\beq
\get(\tau) \ = \ - \frac{1}{3} \sum_{i=1}^3 \frac{\left\langle \Re \, \tr \,
  \left( U(\ltau, \tau) \; g E_i(\tau, \vx) \; U(\tau, 0) \; gE_j(0, \vx)
  \right) \right\rangle}{\left\langle \Re \, \tr \, U(\ltau, 0)
  \right\rangle} \ \cdot
\eeq{get}
Here $U(\tau_1, \tau_2)$ is the gauge link in Euclidean time from
$\tau_1$ to $\tau_2$ at spatial coordinate $\vx$,
$E(\tau, \vx)$ is the color electric field
insertion at Euclidean time $\tau$, $\ltau = 1/T$ is the length of the
Euclidean time direction, $\langle ... \rangle$ indicates
thermal averaging, and an average over the spatial coordinate $\vx$
is implied (see \cite{clm} for a formal derivation). 
  
The spectral function, $\rom$, for the force term is connected 
to $\get(\tau)$ by the integral equation \cite{kapusta} \\
\beq
\get(\tau) \ = \ \int_0^\infty \frac{d \omega}{\pi} \ \rom
\ \frac{\cosh \omega (\tau - \frac{1}{2 T})}{\sinh \frac{\omega}{2 T}} \cdot
\eeq{spectral}
The momentum diffusion coefficient, $\ke$, is then
given by \\
\beq
\ke \ = \ \lim_{\omega \to 0} \ \frac{2 T}{\omega} \ \rom \cdot
\eeq{ke} 
In this work we will use \eqn{spectral}, \eqn{ke} to calculate
the momentum diffusion coefficient $\ke$ for moderately high
temperatures $T \lesssim 3.5 \tc$. In particular, we will be
exploring the temperature dependence of $\ket$.

Note that $\ke$ is the leading order estimate of $\kappa$ in an
$\mqi$ expansion. The $\om(1)$ correction has been explored
\cite{blaine}: modulo some approximations, one can write
\beq
\kappa \approx \ke \; + \; \frac{2}{3} \, \langle v^2 \rangle \, \kb, 
\eeq{om1}
where $\langle v^2 \rangle \; \approx \;
\frac{\textstyle 3 T}{\textstyle M_{\rm kin}}$ is the
thermal velocity squared, and $M_{\rm kin}$ is the kinetic mass,
of the heavy quark.  $\kb$ is the estimate of the diffusion
coefficient one gets by replacing the electric fields with magnetic fields in
\eqn{spectral} and \eqn{ke}. It has been calculated in Ref. \cite{bb}
for the gluonic plasma.

\section{Outline of the calculation}
\label{sec.tech}
We calculated the electric field correlator $\get(\tau)$, \eqn{get}, for
gluonic plasma using lattice discretization and numerical Monte Carlo
techniques. On the lattice, the electric field was discretized,
following \cite{clm}, as
\[ E_i (\vec{x}, \tau) \ = \ U_i (\vec{x}, \tau) \ U_4 (\vec{x}+\hat{i},
\tau) \ - \ U_4 (\vec{x}, \tau) \ U_i (\vec{x}, \tau + 1) \, \cdot \]
Then the lattice discretized $EE$ correlator takes the form
\beq
\gelat (\tau) \; = \; \frac{1}{V} \, \sum_{\vx} \frac{C^i(\tau + 1, \vx) +
  C^i(\tau - 1, \vx) - 2 C^i (\tau, \vx)}{\prod_{x_4=0}^{\ltau} U_4(\vx, x_4)}
\eeq{lcor}
where $C^i(\tau, \vx)$ are Wilson lines at $\vx$ with a hook of length
$\tau$ in the $i$ direction, i.e.,
\[ C^i(\tau, \vx) \; = \; U_i(\vec{x}, 0)  \, \prod_{x_4=0}^{\tau-1}
U_4(\vec{x}+\hat{i}, x_4) \, U_i^\dagger(\vec{x}, \tau) \,
\prod_{x_4=\tau}^{\ltau} U_4(\vec{x}, x_4) \, \cdot \]

We have calculated the correlators $\gelat$ on the lattice at various
temperatures $\lesssim 3.5 \tc$. Equlibrium configurations for a
gluonic plasma were generated at various temperatures by using
lattices with temporal extent $\nt=\frac{\txt 1}{\txt T \, a(\beta)}$,
where $a$ is the lattice spacing, and $\beta = \frac{\txt 6}{\txt \gbsq}$
is the coefficient of the plaquette
term in the Wilson gauge action. The details of the lattices generated
and the number of configurations at each parameter set is given in \tbn{sets}.

\bet
\caption{Summary of runs and statistics.}
\begingroup
\setlength{\tabcolsep}{10pt}
\renewcommand{\arraystretch}{1.5}
\begin{tabular}{ccccccc}
  \hline
  $\beta$ & $\nt$ & $\ns$ & $\ttc$ &
  \# sublattice & \# update & \# conf \\
  \hline
  7.05 & 20 & 64 & 1.50 & 5 & 500 & 1270 \\
  \hline
  7.192 & 24 & 72 & 1.48 & 4 & 2000 & 2032 \\
  \hline
  7.30 & 20 & 64 & 2.03 & 5 & 500 & 1200 \\
  \hline
  7.457 & 24 & 80 & 2.04 & 4 & 500 & 1000 \\
  & 20 & 80 & 2.45 & 6 & 500 & 730 \\
  \hline
  7.634 & 30 & 96 & 2.01 & 5 & 2000 & 640 \\
  & 24 & 96 & 2.51 & 4 & 2000 & 657 \\
  & 20 & 96 & 3.01 & 5 & 2000 & 500 \\
  \hline
  7.78 & 28 & 96 & 2.55 & 7 & 2000 & 678 \\
  & 24 & 96 & 2.98 & 4 & 2000 & 536 \\
  & 20 & 96 & 3.55 & 5 & 2000 & 522 \\
  \hline
  7.909 & 28 & 96 & 2.96 & 7 & 2000 & 1100 \\
  & 24 & 96 & 3.46 & 6 & 2000 & 967 \\
  \hline
\hline
\end{tabular}
\endgroup
\eet{sets}

The spatial extent of the lattices are chosen such that $L T > 3$ and
also the lattice is confined in the spatial direction. 
At various temperatures we have more than one lattice spacings; this
allows us to estimate the discretization error, and to get the
continuum result. Also for various values of the coupling, we have
changed $\nt$ to change the temperature, keeping all the other
parameters of the lattice unchanged. A comparison of the results from
such lattices give us a direct handle on the temperature modification
of $\ke$.

Since we require very accurate correlation functions on lattices with
large temporal extents, we have used the multilevel algorithm \cite{luscher}
in calculating \eqn{lcor}. We follow the implementation of the algorithm
outlined in \cite{prd11}. The number of sublattices
for the multilevel update, and the number of
sublattice updates, are shown in \tbn{sets}, where each update
consisted of (1 heatbath+3 overrelaxation) steps. Typically, a few
parallel streams with independent random number seeds were used at each
parameter sets. After a thermalization run which is many times the
autocorrelation length, $\mathcal{O}(100)$ configurations were
generated from each stream. The total number of configurations
generated at each parameter set is shown in \tbn{sets}.

The temperature scale shown in \tbn{sets} is obtained from the
interpolation formula \cite{biescl}
\beq
\log \frac{r_0}{a} \ = \ \left[ \frac{\beta}{12 \, b_0} \; + \;
  \frac{b_1}{2 \, b_0^2} \, \log \frac{6 b_0}{\beta} \right] \;
\frac{1 \, + \, c_1/\beta \, + \, c_2/\beta^2}{1 \, + \, c_3/\beta \,
  + \, c_4/\beta^2}
\eeq{biescl}
where $b_0 \, = \, 11/(4 \pi)^2$ and $b_1 \, = \, 102/(4 \pi)^4$. The fit
parameters $c_i$ are 
\beq
c_{\{1,2,3,4\}} \ = \ \{ -8.9664, \, 19.21, \, -5.25217, \, 0.606828 \}
\eeq{fit}
and $r_0 T_c$ = 0.7457 \cite{biescl}. Other ways of determining the
temperature gives slightly different values: e.g., using the formula of
Ref. \cite{ehk} leads to a temperature which differs by
$\sim$ 1-1.5 \% at the higher $\beta$ values of \tbn{sets}. So we will
effectively round off the temperature and, e.g., treat 3.46 $\tc$ and
3.55 $\tc$ in \tbn{sets} as $\sim 3.5 \tc$.

\section{Analysis of the correlators and extraction of $\kt$}
\label{sec.results}

\subsection{Discretization effect in $\gelat$}
\label{sec.cutoff}

The $EE$ correlation functions $\get(\tau)$ are ultraviolet finite. The
bare correlators $\gelat(\tau)$ require only finite renormalization:
\beq
\gern (\tau) \; = \; \zee(a(\beta)) \, \gelat (\tau) \, \cdot
\eeq{renorm}
The renormalization coefficient $\zee(a)$ has been determined at one loop
level in \cite{1lp}:
\beq
\zee \ = \ 1 \, + \, \frac{2 \gbsq C_f}{3}
\; P_1 \ \approx \ 1 \, + \, 0.1377 \gbsq
\eeq{1lp}
where the lattice bare coupling $\gbsq = \frac{\txt 6}{\txt \beta}$,
$C_f=\frac{\txt 4}{\txt 3}$,
and $P_1 = \int_{-\pi}^\pi \, \frac{\textstyle d^4k}{\textstyle (2 \pi)^4} \,
\frac{\textstyle 1}{\textstyle K^2} \, \approx 0.15493$.

After the renormalization, the correlator still shows cutoff effect,
especially at short distances. A major part of this cutoff effect at
short distances can be taken into account by a consideration of the
discretization effect in the leading order. In leading order, the EE
correlator takes the form \cite{francis}
\begin{eqnarray}
  \get(\tau, LO) &=& g^2 \, \cf \; \gfr(\tau), \\
  \gcont(\tau) &=& \pi^2 T^4 \; \cosec^2(\pi \tau T) \ \left(
  \cot^2(\pi \tau T) \, + \, \frac{1}{3} \right) \label{freecont} \\
  \glat(\tau) &=& \frac{1}{3 a^4} \ 
  \dtk(k) \frac{e^{\bar{k} (\nt - t)} \, + \, e^{\bar{k}
      t}}{e^{\bar{k} \nt} \, - \, 1} \ \frac{\tilde{k}^2}{\sinh
    \bar{k}} \label{freelat}
\end{eqnarray}
where the integration is over the Brillouin zone $(-\pi,\pi)$ and
\beq
\frac{\bar{q}}{2} \; = \; \arcsinh \frac{\tilde{q}}{2} \; , \qquad
\tilde{q}^2 \; = \; \sum_{i=1}^3 \; 4 \, \sin^2 \frac{q_i}{2} \ \cdot
\eeq{latmom}
A major part of the discretization effect can be accounted for by a
comparison of \eqn{freecont} with \eqn{freelat}: in particular, by
defining an {\em improved distance} $\ti$ through \cite{sommer,francis}
\beq
\glat(\ti) \; = \; \gcont(\tau) \, \cdot
\eeq{tim}
We have $\gelat(\tau)$ at multiple lattice spacings at each temperature.
As noted in \cite{bielefeld,tum1} before, we found that the use of $\ti$,
\eqn{tim}, reduces considerably the short distance discretization
effect in $\get(\tau)$.  In what follows, we have used $\ti$ to denote the
distance scale for $\gern$. 

At $T/\tc$ = 2, 2.5 and 3 we have three lattice spacings
each. Using these correlators, we find the continuum extrapolated
correlators at distances corresponding to $\ti$ for the finest
lattice. We extrapolate $\ggfr$ to $a \to 0$, where
$\tau$ takes the values $\ti$ of the finest lattice at each
temperature. The details of the method are presented in \apx{details}. 
The correlators $\ggfr$ for the different discretized lattices, and their
continuum extrapolated value, are
shown in \fgn{cont}. The extrapolated ratio is now multiplied by
$\gcont(\tau)$ to get the continuum extrapolated correlator. The calculation
is done through a bootstrap analysis. For the bootstrap, the data is
first blocked in blocks of size at least 2-3 times the autocorrelation time. 

\bef
\centerline{\includegraphics[scale=0.5]{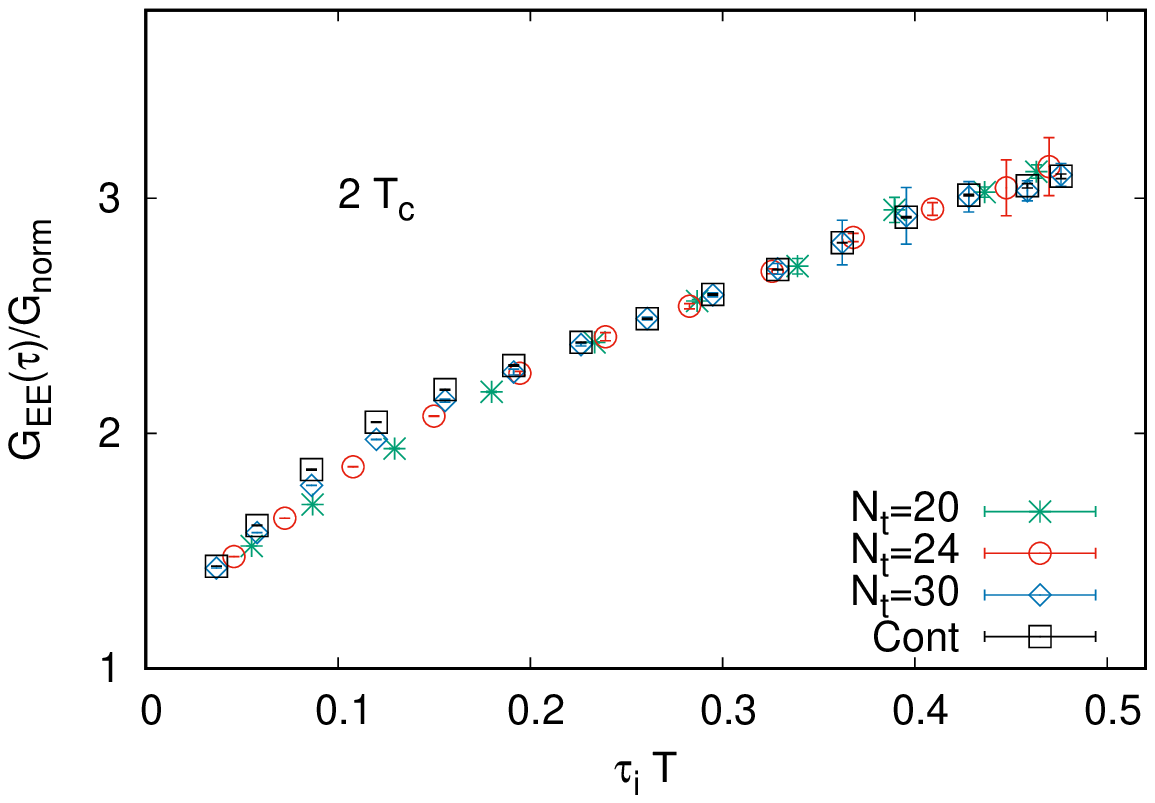}\includegraphics[scale=0.5]{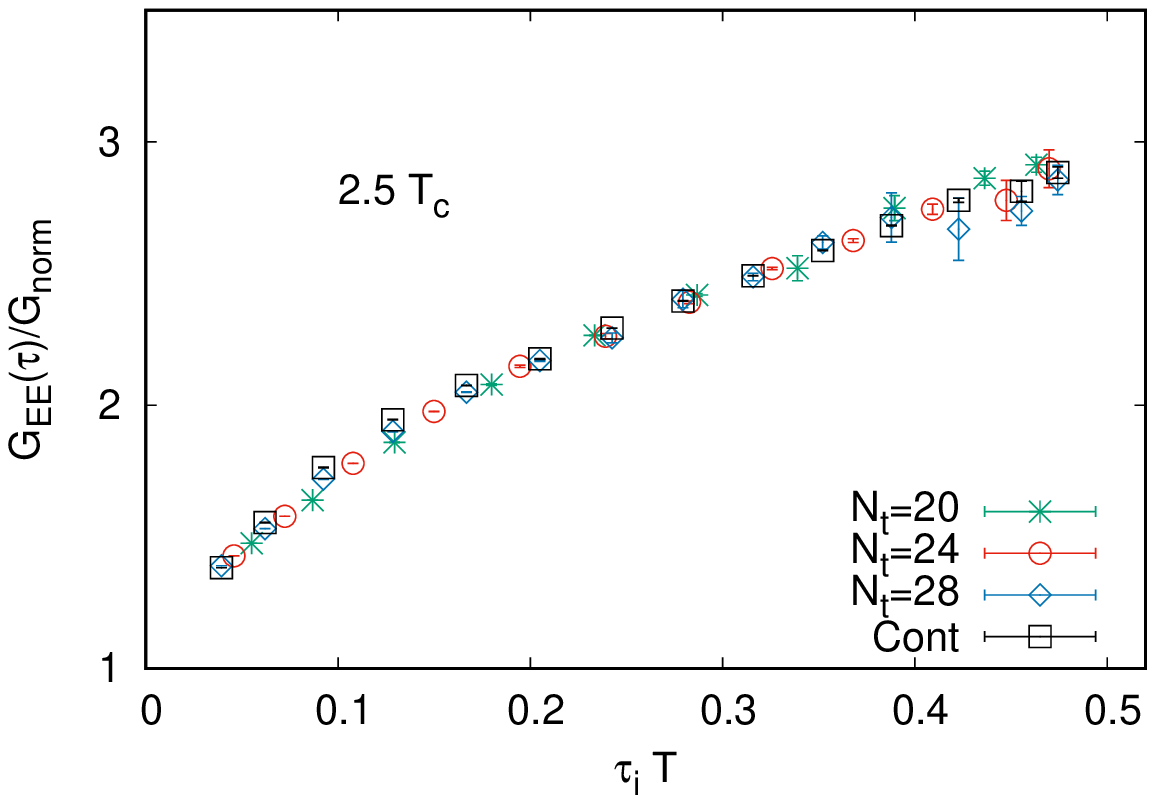}\includegraphics[scale=0.5]{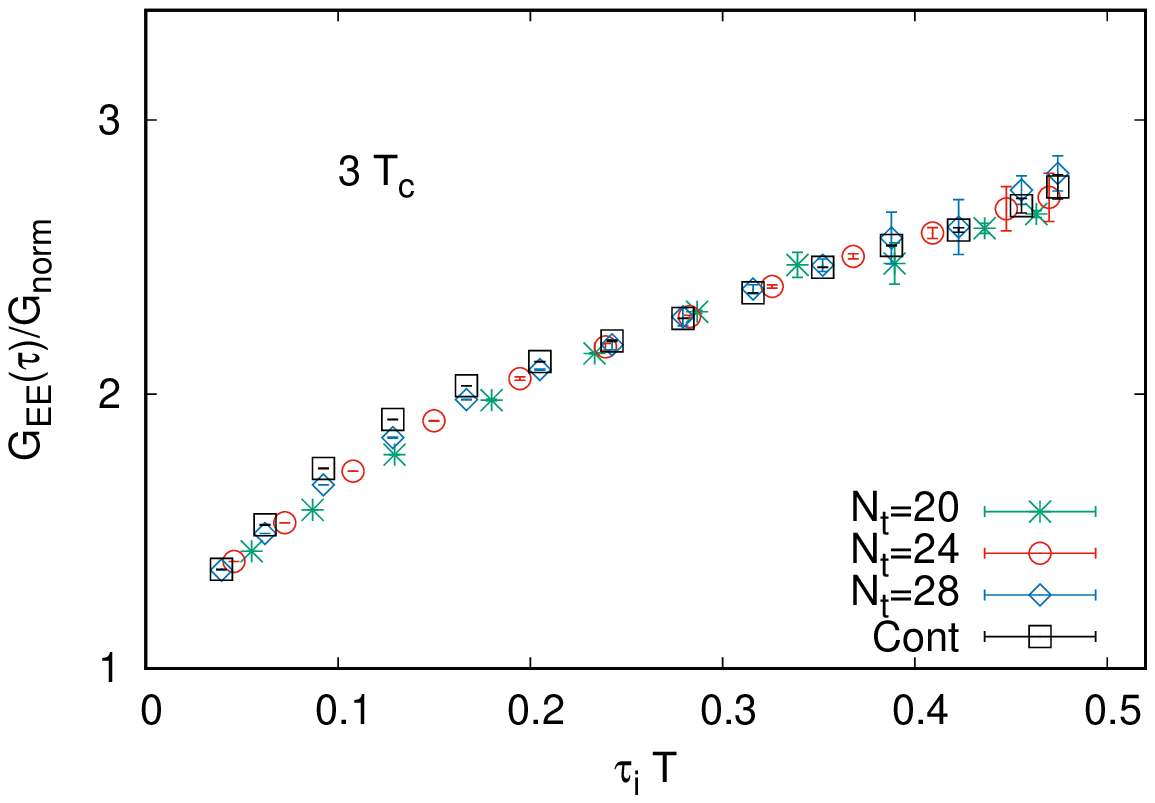}}
\caption{Continuum extrapolation of the correlator ratio $\ggfr$,
  at (left) 2 $\tc$, (middle) 2.5 $\tc$, and (right) 3 $\tc$.}
\eef{cont}

It is interesting to compare the continuum extrapolated correlator
with perturbation theory. $\get$ has been calculated in perturbation
theory to NLO in Ref. \cite{burnier}; in \fgn{pertcorr} we compare the
perturbative results of Ref. \cite{burnier} with our nonperturbatively
determined correlator \cite{lonlo}. In Ref. \cite{burnier} the scale for the
running coupling has been set at 
\beq
\mu_{\rm opt} \, \approx \, \max \, [7.57 \, \omega, \; 6.74 \, T ] 
\eeq{pms}
following the {\em principle of minimal sensitivity}.
The LO and NLO bands in \fgn{pertcorr} are obtained by varying
$\mu \, \in \, [0.5,2] \, \mu_{\rm opt}$.
This way of scale setting leads to a good agreement between 
the LO and the NLO calculation; but as \fgn{pertcorr} shows,
the perturbative estimates are very different from the nonperturbative results.
We also show the LO results obtained by setting the scale 
in an intuitive way \cite{bielefeld}:
\beq
\mufit \, = \, \max \, [\omega, \pi T] \, \cdot
\eeq{mufit}
The band is obtained by varying this scale by a factor $[0.5,2.0]$
as before. As \fgn{pertcorr} shows, the LO curve captures the main features
of the nonperturbative result. However, the good agreement of perturbation
theory with the lattice result in this case is
misleading, as the NLO result changes strongly from the LO result
and the lattice result. Guided by \fgn{pertcorr}, we will  
use the LO spectral function evaluated at the scale $\mufit$ for modelling
the ultraviolet part of the spectral function.  

\bef
\centerline{\includegraphics[scale=0.8]{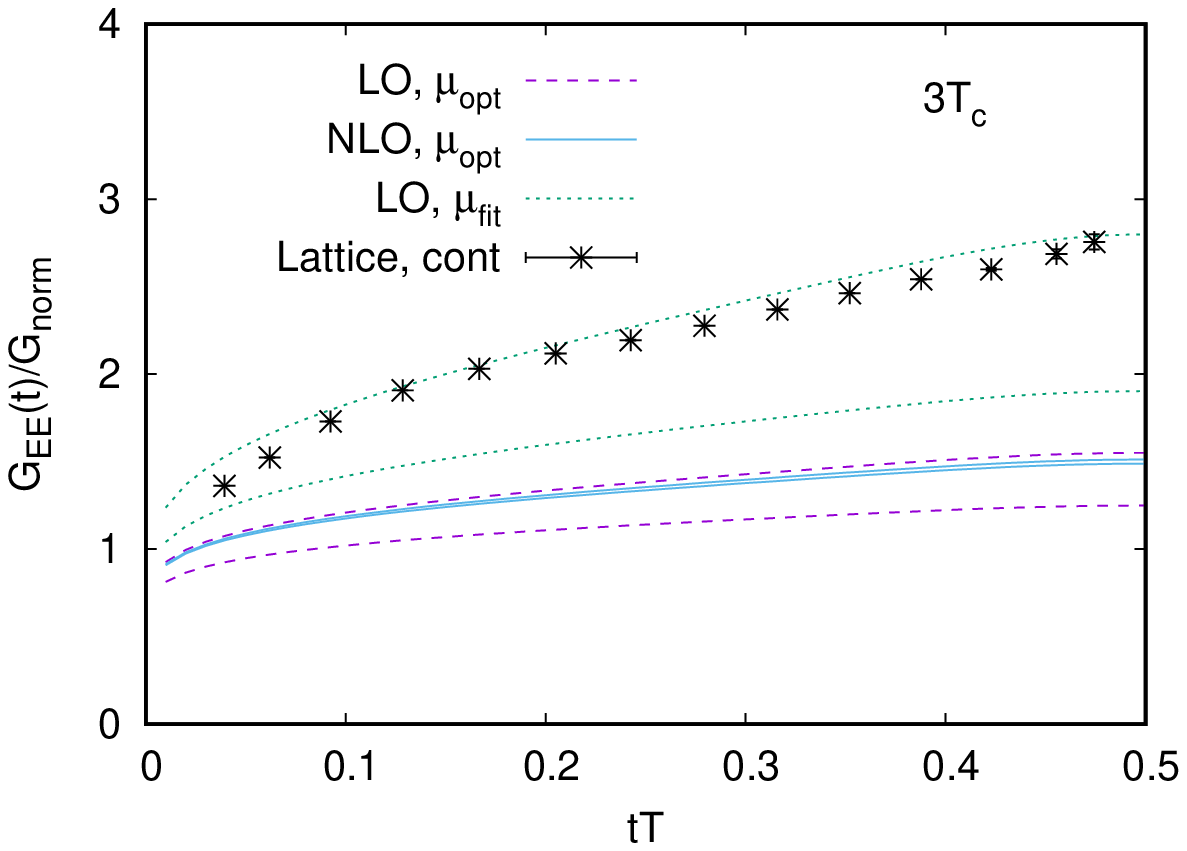}}
\caption{A comparison of the nonperturbatively obtained correlator
  $\ggfr$ at $T=3 T_c$ with the results of perturbation theory \cite{burnier}.
  The LO and NLO results use an optimized scale (\eqn{pms}). 
  Also shown is the LO result with the scale
  $\mufit$ (\eqn{mufit}). In each case, the bands for the
  perturbation theory are obtained by varying the scale by a factor of
  two in each direction from the scale mentioned above.}
\eef{pertcorr}

\subsection{Extraction of $\kappa$ from the correlators}
\label{sec.kbt}
A direct inversion of \eqn{spectral} to get $\rom$ is
very difficult. Instead, to get an estimate of what kind of $\ke$ is
consistent with the $\get$ obtained, we have used some simple models for
$\rom$. Our models, and the analysis strategy, are similar to what was
followed in Ref. \cite{bb} for $\gbt$, which, in turn, was influenced by
earlier works \cite{prd11,bielefeld} on $\get(\tau)$. The ultraviolet
and the infrared parts of $\rom$ are modelled with the simple
forms
\beq
\puv \; = \; \frac{g^2(\mufit) \, C_f \, \omega^3}{6 \pi}, \qquad
\pir \; = \; \ke \, \omega \, ,
\eeq{rmod}
where $\mufit$ is defined in \eqn{mufit}. $\pir$ is the simplest form
capturing the dissipative behavior of $\ke$. The correlator \eqn{get} does
not have a transport peak, and is expected to have a smooth linear behavior
in the infrared \cite{clm,burnier}, motivating $\pir$. $\puv$ is the known
leading order form of the spectral function and the scale choice is motivated
by \fgn{pertcorr}. The NLO spectral function is known \cite{burnier}
but, as \fgn{pertcorr} shows, it is not clear that it will capture the
ultraviolet behavior better except at very high $\omega$.

While both $\puv$ and $\pir$ are well-motivated, not much is known {\em
  a-priori} about the form of the spectral function in the intermediate
$\omega$ regime. An ansatz, that allows 
$\rom$ to continuously change from $\puv$ to $\pir$, is
\beq
\rom \ = \ \max \left[ c \, \puv , \; \pir \right]
\eeq{formc}
where we have introduced a parameter $c$ to take into account the
uncertainty due to the scale choice and the use of the leading order
form for $\puv$. $c$ is treated as a fit parameter.
The best fit values we obtained for $c$ are close to 1, in the range 1-1.2.

A more smooth form of connecting $\puv$ with $\pir$ is
\beq
\rom \ = \ \left[\sqrt{(c \, \puv)^2 \; + \; \pir^2}
  \right] \cdot
\eeq{formb}
The form of \eqn{formb} has been argued to be theoretically better justified
in \cite{bielefeld}, \cite{gflow}. Here again, the fit parameter $c \sim 1$
is introduced to account for the uncertainty in $\puv$.
In our analysis we have treated \eqn{formc} and \eqn{formb} at par.

Instead of introducing a fit parameter $c$ as above, ref. \cite{bielefeld} has
suggested parametrizing the difference between the above forms (with $c$=1) and
$\rom$ in a sine expansion:
\[
\left(1 \, + \, \sum_n \, c_n \, \sin(\pi n y) \right), \qquad y \, = \,
\frac{x}{1+x}, \ \ x \, = \, \log \left( 1 \, + \, \frac{\omega}{\pi T} \right)
\, \cdot
\]
For the fit range we used, we found that one term in the expansion sufficed
to fit our correlator. Therefore we have also tried the fit forms
\begin{eqnarray}
  \rom &=& \left( 1 \, + \, c_1 \, \sin \, \pi y \right) \
  \left[\sqrt{\puv^2 \; + \; \pir^2}
    \right] \, ; \label{fourierb} \\
  &=& \left( 1 \, + \, c_1 \, \sin \, \pi y \right) \
  \max \left[ \puv , \; \pir \right] \, . \label{fourierc} 
\end{eqnarray}
In all our fits we have found $c_1$ to be small, $\in [0.02,0.12]$.

We perform the whole analysis for each of the model forms \eqn{formc},
\eqn{formb}, \eqn{fourierb} and \eqn{fourierc}
in a bootstrap framework. Our final estimates of $\ke$ are
 shown in \tbn{ket} and in \fgn{ket}.  The details of the analysis can
 be found in \apx{details}, where the estimates for each model are
 given in \tbn{kbtforms} and \fgn{ke-fits}. Our estimates in \tbn{ket}
 and \fgn{ket} include the entire bands for \eqn{formb}, \eqn{formc}
 and the central values for \eqn{fourierb}, \eqn{fourierc}.

\bef
\centerline{\includegraphics[scale=0.7]{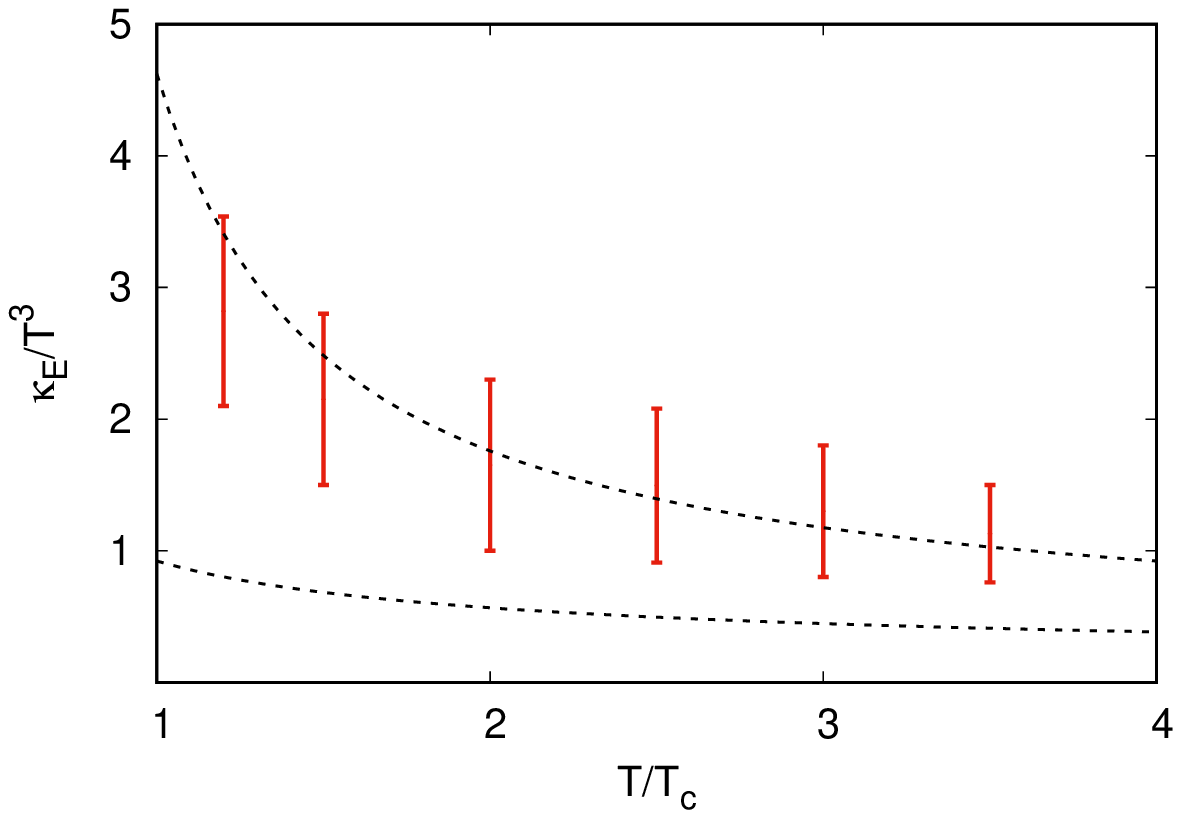}
  \includegraphics[scale=0.7]{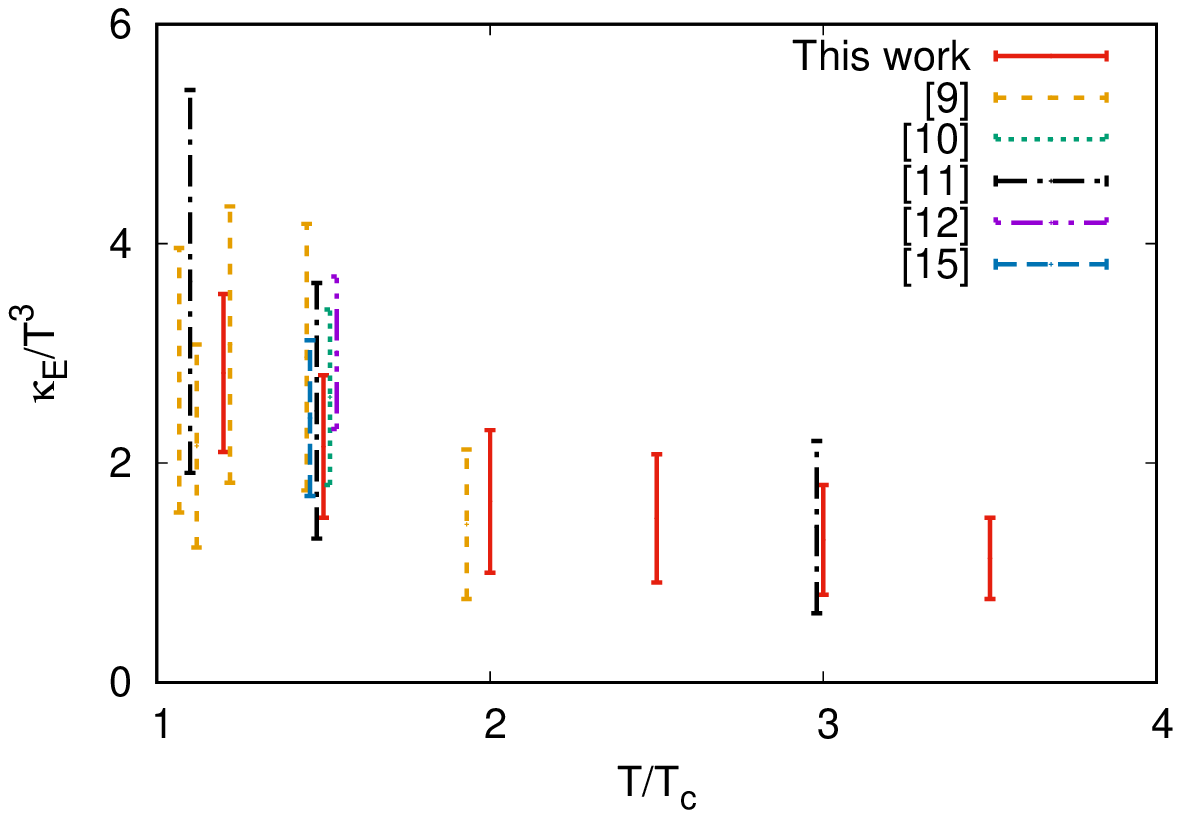}}
\caption{(Left) Our estimates for the range of $\ket$ in the temperature
  range $\lesssim 3.5 \tc$. Also shown (dotted lines) is the NLO
  perturbation theory estimate \eqn{nlo} \cite{cm}; the band
  corresponds to varying the
  scale of the coupling $g^2(\mu)$ in the range $\mu \in [\pi T, 4 \pi T]$.
  (Right) A survey of other existing lattice results for $\ke$ in
  gluonic plasma in the 1-4 $\tc$ temperature range. For visual clarity,
  points at 1.5 $\tc$ and 3 $\tc$ have been slightly shifted horizontally.}
\eef{ket}

\bet
\caption{Temperature dependence of $\ket$}
  \begingroup
\setlength{\tabcolsep}{10pt}
\renewcommand{\arraystretch}{1.5}
\begin{center}
  \begin{tabular}{c|cccccc}
    \hline
  $T/\tc$ & 1.2 & 1.5 & 2.0 & 2.5 & 3.0 & 3.5 \\
  $\ket$ & 2.1 - 3.5 & 1.5 - 2.8 & 1.0 - 2.3 & 0.9 - 2.1 &
    0.8 - 1.8 & 0.75 - 1.5 \\
    \hline
\end{tabular} \end{center}
\endgroup
\eet{ket}

There are other estimates of $\ket$ for a gluonic plasma from the
lattice. While Ref. \cite{prd11} studied the temperature range
close to $\tc$, a detailed study at 1.5 $\tc$ was performed in
Ref. \cite{bielefeld}. A broad temperature range was studied in
Ref. \cite{tum1}, with the main focus being very high
temperatures. While the analysis techniques, in particular the
spectral function models, vary, all these references used the multilevel
algorithm and perturbative renormalization constants. Recently, Refs.
\cite{gflow} and \cite{tum2} have used gradient flow
\cite{luscher-flow} to get the renormalized $EE$ correlators at 1.5
$\tc$. We compare these studies with ours in the right panel of
\fgn{ket}.  Within the uncertainties of our and other studies,
our results agree very well with the other studies.

\section{Summary and Discussion}
\label{sec.summary}
In this paper we have studied the electric field correlator,
\eqn{get}, in a thermally equilibriated gluonic plasma at moderately
high temperatures $T \lesssim 3.5 \tc$. We investigated in detail the
cutoff dependence of the correlators (\fgn{cont}). With a simple set
of models for the $EE$ spectral function $\rom$, we then estimated the
static quark momentum diffusion coefficient $\ke$. The results are
shown in \fgn{ket} and in \tbn{ket}.

$\ket$ has been calculated to NLO in perturbation theory in \cite{cm}.
For SU(3) gluonic plasma, the NLO result is
\beq
\ket \ = \ \frac{g^4 C_F}{6 \pi} \, T^3 \, \left[ \log \frac{2
    T}{\md} \, + \, \xi \, + \, C \, g \right]
\eeq{nlo}
where $C_F=4/3, \, \xi \, \approx \, -0.64718, \ C \approx 2.3302$
and $\md = g T$ in LO perturbation theory.
This NLO result is shown in \fgn{ket} by the band bordered by the dotted
lines; the band corresponds to
evaluating $g^2$ at the scales $\mu \in [\pi T, 4 \pi T]$. The NLO
results explain the data quite well. Note, however, that perturbation
theory is inherently unstable here: the LO result is an
order-of-magnitude smaller than NLO. In fact, if we omit the
$\mathcal{O}(g)$ term in \eqn{nlo}, we will get a negative value for $\ket$ 
in our temperature range \cite{burnier}. The agreement of the NLO
result with the nonperturbative results may indicate that the
corrections beyond NLO are small.

It is of interest to compare our results on the temperature dependence of
$\ke$ with some other theoretical calculations. The estimate in Ref. \cite{ct}
is for $\mathcal{N}=4$ supersymmetric Yang-Mills theory, which is scale
invariant. Clearly, $\ket$ is temperature independent in such a theory.
A different AdS-CFT based approach has been taken in Ref. \cite{andreev}, where
the drag term has been connected to the spatial string tension, $\sigma_s$.
Eqn. (\ref{fd}) then connects $\ke$ to $\sigma_s$. While the estimate of
$\ket$ obtained in Ref. \cite{andreev} this way is close to our estimates, it
provides a somewhat milder temperature dependence at higher temperatures.

While it is the momentum diffusion coefficient that enters the
equations of Langevin dynamics and is of interest for the
phenomenology of heavy quark thermalization, it has been the
convention to quote the transport coefficient as the position space
diffusion coefficient $D_s$. In particular, the combination
\beq
\ds \; = \; \frac{4 \pi}{\kt}
\eeq{ds}
is usually quoted. In the left panel of \fgn{ds} we plot $\ds$ as obtained
form our results of $\ke$ using \eqn{ds}.

\bef
\centerline{\includegraphics[scale=0.8]{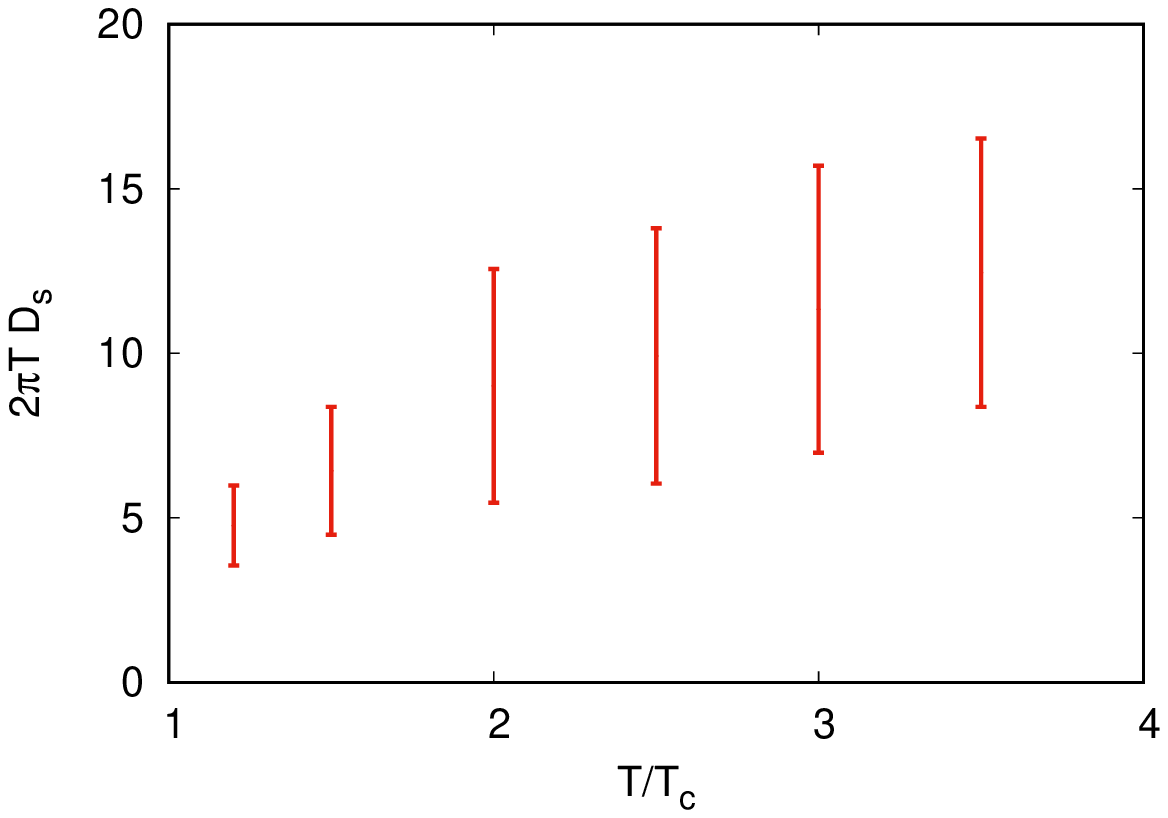}}
\caption{An estimation of the static quark diffusion
  coefficient, using \eqn{ds} and \tbn{ket}.}
\eef{ds}

For phenomenological studies, one is interested in estimates of $\kappa$
for charm and bottom, rather than for the infinitely massive quarks. Using
\eqn{om1} \cite{blaine} one can provide such an estimate \cite{bb}. In
\fgn{dscb} we show the separate estimates for $\dsc$ and $\dsb$, obtained
using \eqn{om1} and \eqn{ds}, and the $\ke$ values of \tbn{ket}.
The estimates of $\kb$ are from
ref. \cite{bb}, supplemented by a calculation at 3 $\tc$, following
exactly the same analysis techniques as in Ref. \cite{bb}.
The estimates we get for $\dsc$ and $\dsb$ are shown in \fgn{dscb}.
The details can be found in \apx{details}.

\bef
\centerline{\includegraphics[scale=0.8]{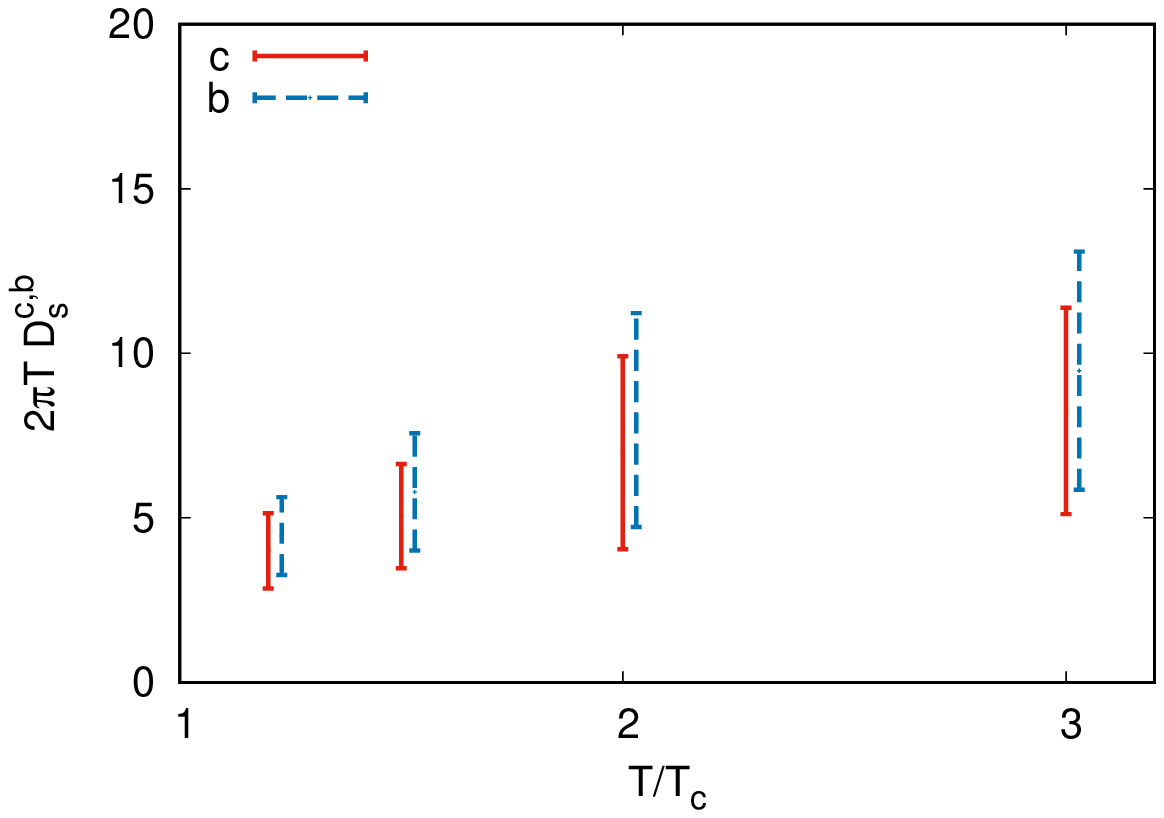}}
\caption{Estimate of $\dsc$ and $\dsb$, the spatial diffusion coefficients
  for the charm and bottom quarks, using \eqn{om1} and \eqn{ds}. See text.
  For visual clarity, the points for $\dsb$ have been slightly shifted
  horizontally.}
\eef{dscb}

$\dscb$ show a rising trend with temperature. The temperature
dependence of $\ds$ is of great interest to phenomenological studies
\cite{cao,duke,das}. In particular, in Ref. \cite{duke}, using
a parametric temperature dependence
\beq
\ds \sim \alpha \; + \; \gamma \, \left(\frac{\txt T}{\txt \tc} -1 \right),
\eeq{ds-linfit}
$\ds$ was estimated from the experimental data for $D$ meson using a Bayesian
analysis. They quote the central values $\left(\alpha, \, \gamma \right)
\, \sim \, \left( 1.9, \, 3.0 \right)$, with $\alpha \sim 1-3$ being the
$5-95$ percentile band \cite{duke}. While our study is for quenched QCD,
it is still interesting to check if the temperature dependences of 
$\dsc$ and $\dsb$ shown in \fgn{dscb} are consistent with the simple
parametrization of \eqn{ds-linfit}. The answer is ``yes'' 
(admittedly, aided by the large uncertainties in our measurements),
with $\left(\alpha, \, \gamma \right)$ = $\left( 3.61(30), \, 2.57(43) \right)$
for charm and $\left( 3.99(35), \, 3.08(54) \right)$ for bottom, respectively
\cite{caution}. For the static $D_s$, using the same parametrization
we obtained $\alpha$ = 4.27(29) and
$\gamma$ = 3.60(33). We also tried doing this linear fit
for $\ds$ from each of the models of \scn{kbt}. The results can be seen
in \fgn{dsforms} and \tbn{ds-linfit} in \apx{details}. All of the
model spectral functions indicate a positive slope of $\ds$ with temperature.
We emphasize that \eqn{ds-linfit} is a purely phenomenological fit:
the temperature dependence of $D_s$ is of course more complicated,
e.g., \eqn{nlo}.

\section{Acknowledgements}
We thank Mikko Laine for providing the perturbative curves in
\fgn{pertcorr}, and for discussions.  The computations presented in
this paper were performed on the clusters of the Department of
Theoretical Physics, TIFR, and on the ILGTI computing facilities of
IACS and TIFR.  We would like to thank Ajay Salve and Kapil Ghadiali
for technical assistance.  S.D.\ acknowledges support of the
Department of Atomic Energy, Government of India, under Project
Identification No.\ RTI 4002.

\appendix
\section{Some details of the numerical analysis}
\label{sec.details}
Here we provide additional details of our numerical analysis in \scn{results}.

We do a detailed calculation at temperatures $\ttc$ = 2, 2.5 and 3.
At these temperatures we have correlators from three lattice spacings.
We have estimated the continuum correlator from them, and calculated $\ke$
from it. Our whole analysis is done in a bootstrap formalism.
To get the continuum correlator, the
average correlator for each bootstrap sample is B-spline interpolated,
and the value of the correlation function at distances corresponding
to the $\ti$ values of the finest lattice are obtained. Note that this
typically involves an {\em extrapolation} at the smallest distance,
but this is not a concern as this distance is not used in the fits.
A very slight extrapolation is also required at the largest ($\tau \sim 1/2T$)
distance point, but it is a very small extrapolation and we do not
expect this to be a problem.

At short distances $\ti T \lesssim 0.15$ we see a clear discretization
effect, which is approximately linear in $a^2$; we fit to a linear
form to get the continuum correlator. For larger distances $\ti T >
0.25$ the correlators do not show a clear discretization effect. In
particular, for correlators at large distances $\ti T \gtrsim 0.3$
we found a constant fit to be more reasonable. We show examples of our
continuum extrapolation at some representative distances in
\fgn{extrapol}. Note that we have also carried out the analysis with
linear extrapolation at all distances; the $\kappa$ values obtained
agree within errorbar.

To extract $\ke$ from the continuum correlators using \eqn{spectral},
we have used the fit forms discussed in \scn{kbt}, and done a standard
$\chi^2$ fit. The [16,84] percentile band of the bootstrap estimators
of $\ke$ is treated as the 1-$\sigma$ errorband. 
The results for the various fit forms are shown in
\fgn{ke-fits}. Typically we get a good $\chi^2$ by taking the whole
range except the two shortest distance points. We have, however, also
varied $\tmin$. The results shown in \fgn{ke-fits} include the
variation with fit range, and any difference due to using linear vs
constant extrapolation at large separations in \fgn{extrapol}.

As mentioned in \scn{kbt}, we have also fitted the correlators from
the individual lattices to the forms of \scn{kbt}. For this we have
used $\tmin \sim 0.25/T$, where the discretization effect
on the correlators is small. $\tmin$ is further varied within a small
range. The bands shown in \fgn{ke-fits} include the spread due to such a
variation.

In \tbn{kbtforms} we show the final results for $\ket$ using the
different fit forms. The error estimate is conservative, covering the
1$\sigma$ interval obtaned from the continuum correlator and the
correlators from lattices with $N_t \ge 24$. 

\bet
\caption{Results for $\ket$ from the different fit forms of
  \scn{kbt}.}
\begingroup
\setlength{\tabcolsep}{10pt}
\renewcommand{\arraystretch}{1.5}
\begin{center}
  \begin{tabular}{ccccc}
    \hline
    $T/\tc$ & \eqn{formc} & \eqn{formb} & \eqn{fourierc} &
    \eqn{fourierb} \\
    \hline
    1.2 & 2.16-2.80 & 2.44-3.54 & 1.80-2.50 & 2.34-3.14 \\
    1.5 & 1.74-2.16 & 1.62-2.80 & 1.25-1.73 & 1.55-2.27 \\
    2   & 1.05-1.60 & 1.48-2.30 & 0.77-1.42 & 1.04-1.82 \\
    2.5 & 0.91-1.77 & 1.17-2.08 & 0.70-1.59 & 0.97-1.86 \\
    3   & 0.87-1.48 & 1.04-1.80 & 0.60-1.30 & 0.83-1.56 \\
    3.5 & 0.76-1.14 & 1.01-1.50 & 0.62-1.02 & 0.96-1.33 \\
    \hline
\end{tabular} \end{center}
\endgroup
\eet{kbtforms}

\tbn{kbtforms} also includes two temperatures where we have
only two lattice spacings each, and 1.2 $\tc$ where we have reanalyzed
the correlators on $N_t$=24 lattices calculated in Ref. \cite{bb}.
In these cases we have only fitted the individual
lattices. The rest of the discussion is the same as above. The final
result in these cases is taken from the $N_t$=24 lattices.

For the final result for $\ke$ shown in \fgn{ket}, we have treated the
fit forms \eqn{formc} and \eqn{formb} at par, and conservatively
quoted an error band that includes the bands for \eqn{formc} and
\eqn{formb} in \tbn{kbtforms} and the central values of the bands for
\eqn{formc} and \eqn{formb}. These results are also shown in
\tbn{ket}.

\bef
\centerline{\includegraphics[scale=0.5]{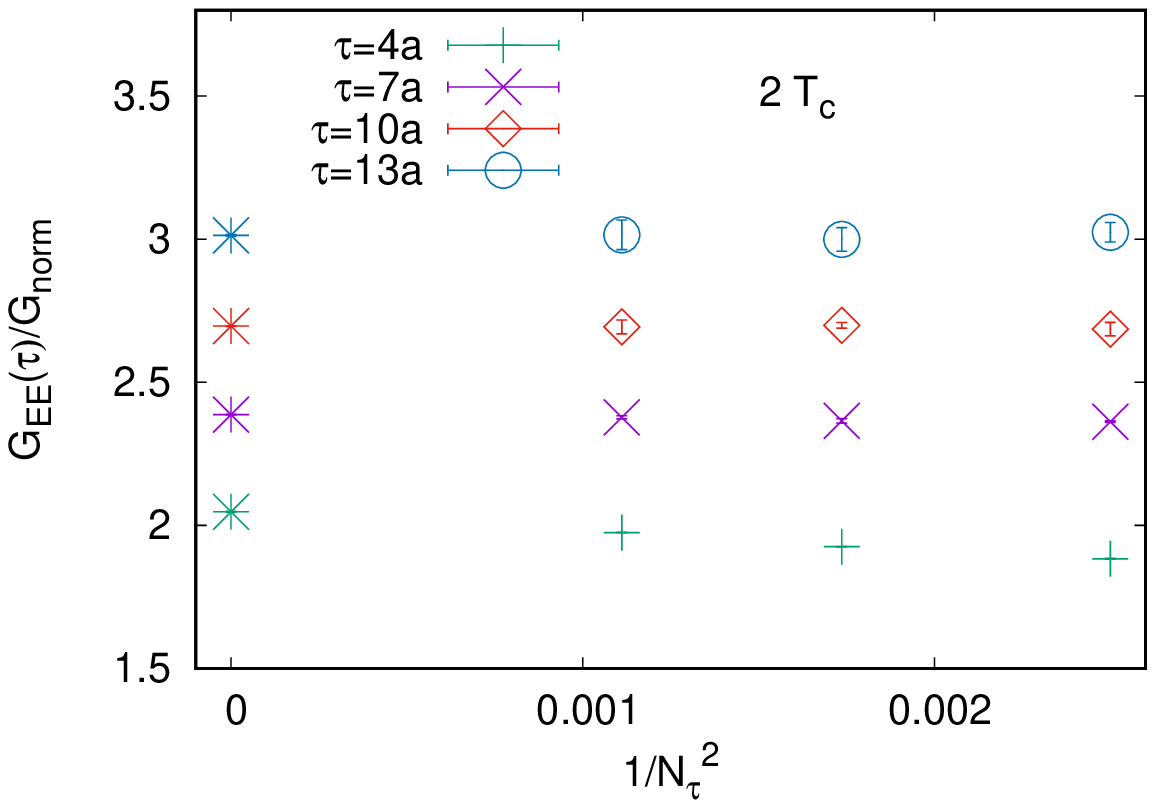}
  \includegraphics[scale=0.5]{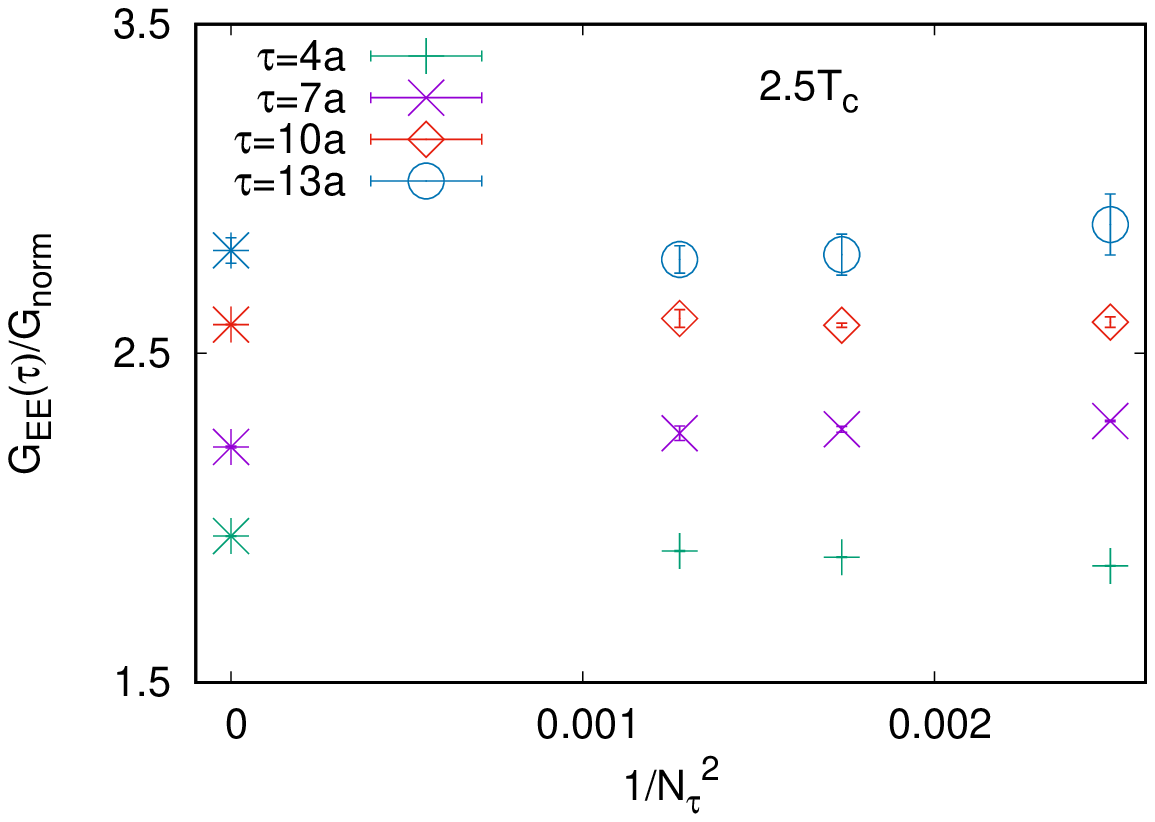}
  \includegraphics[scale=0.5]{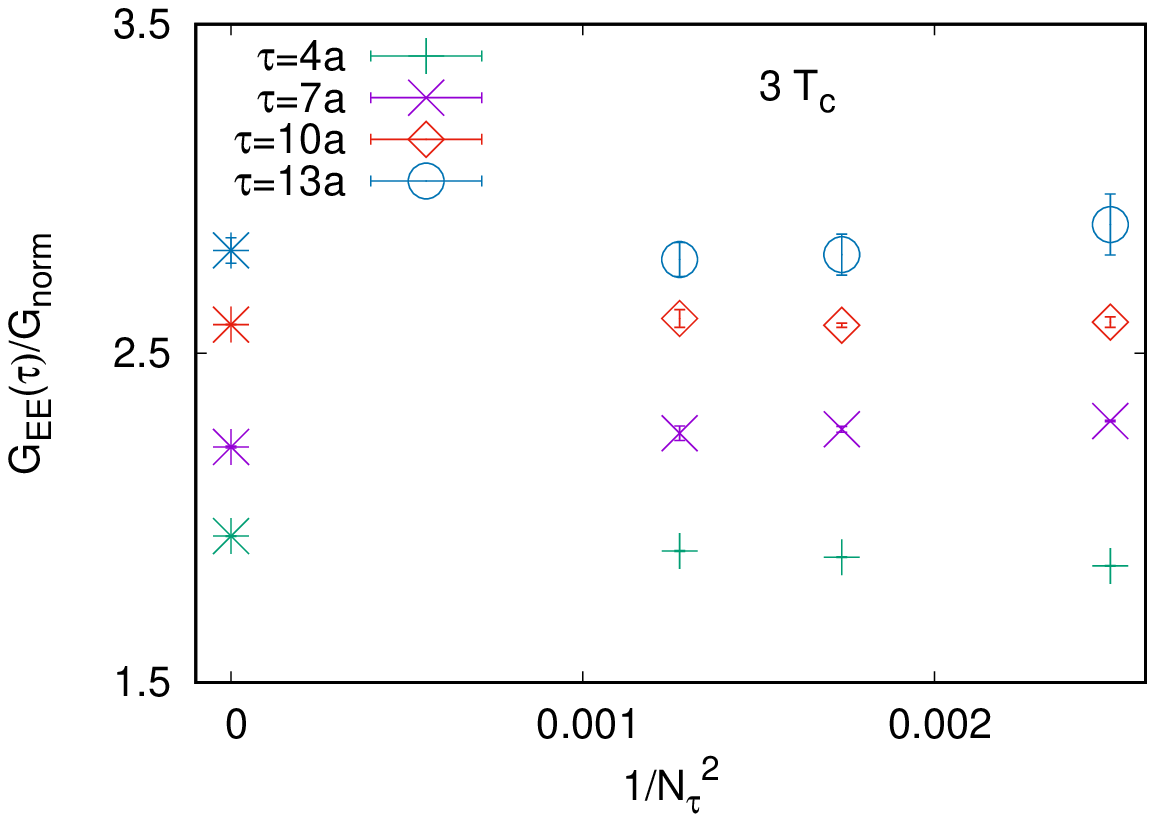}}
\caption{Illustration of the continuum extrapolation of the correlator
  normalized by $\gfr$. (Left) 2 $\tc$, (middle) 2.5 $\tc$ and(right) 3 $\tc$.
  }
\eef{extrapol}

\bef[t]
\centerline{\includegraphics[scale=0.5]{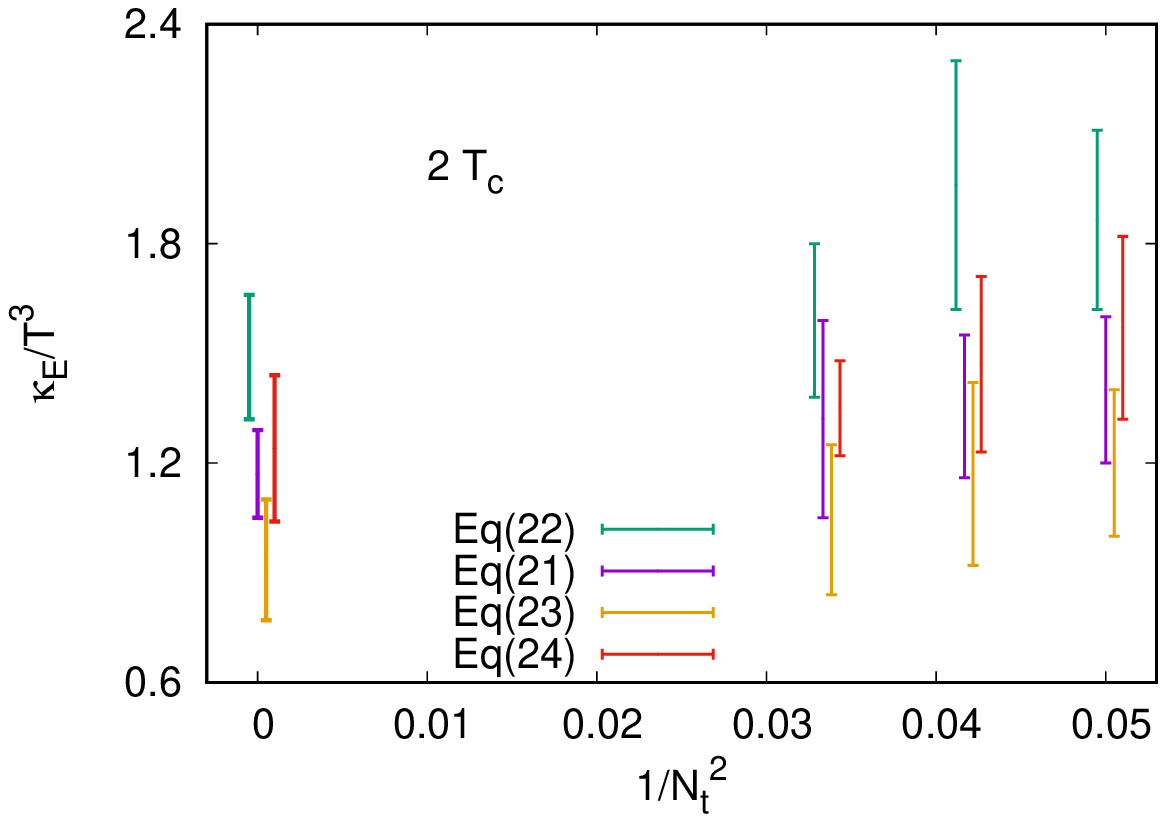}
  \includegraphics[scale=0.5]{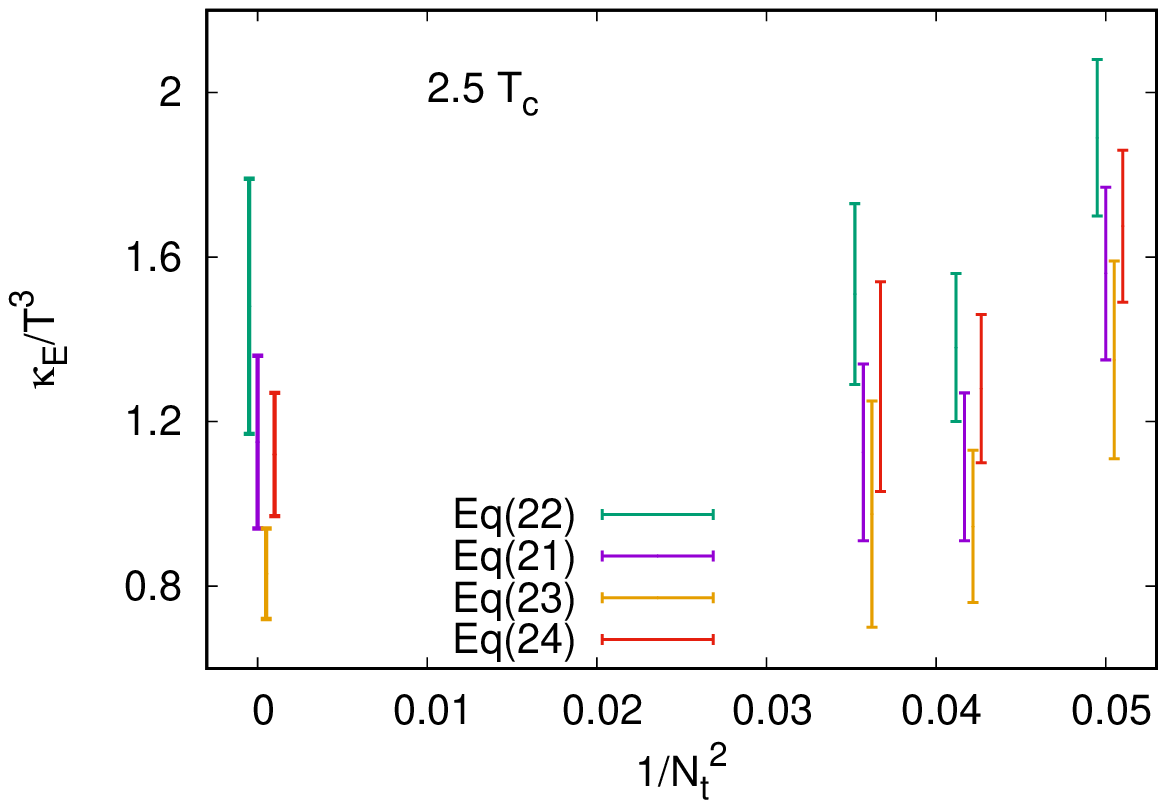}\includegraphics[scale=0.5]{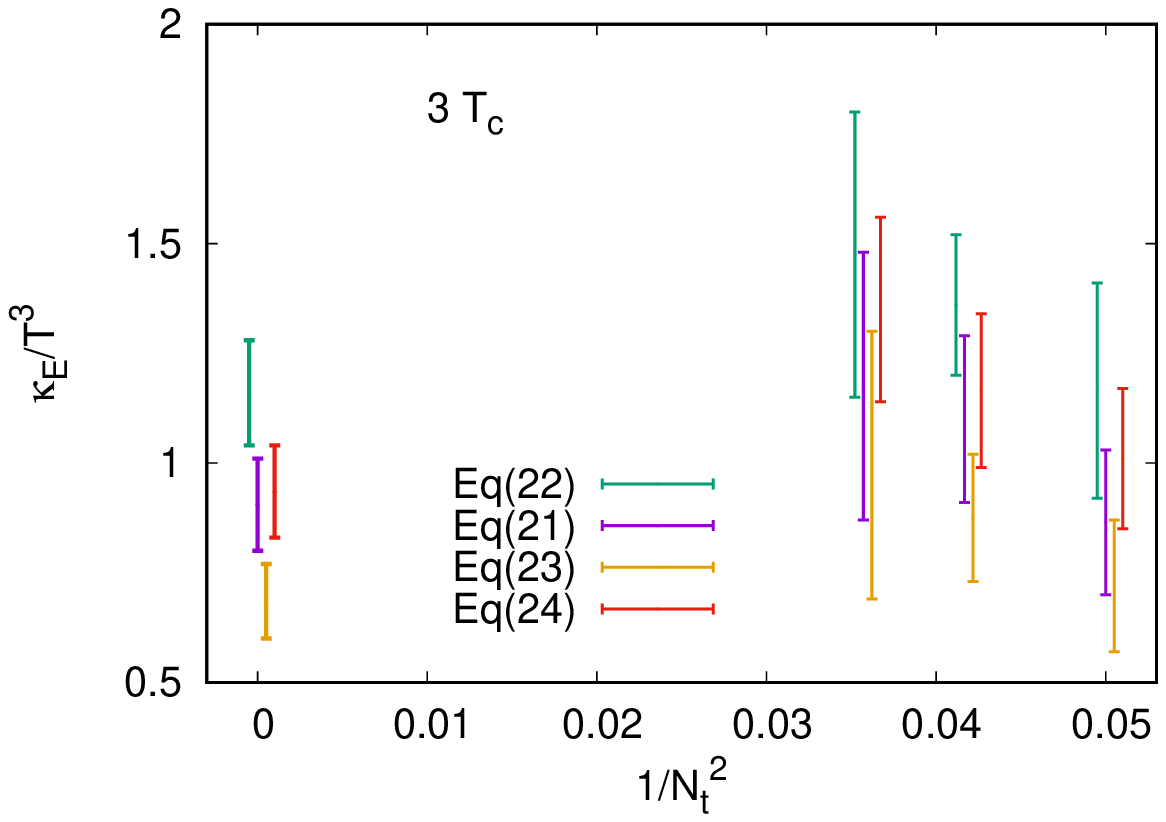}}
\caption{Results for $\ket$ obtained at $T/\tc$ = 2 (left), 2.5
  (middle) and 3 (right). Besides the continuum results, the results
  obtained from fitting individual lattices is also shown. See the
  text for details of the error band.}
\eef{ke-fits}

Using \tbn{kbtforms} and \eqn{ds} we can also make separate estimates for
$\ds$ for each form of the model $\rom$ in \scn{kbt}. This is shown in 
\fgn{dsforms}. The linearly rising behavior of each of these forms can
then be separately fitted to the linear fit form \eqn{ds-linfit}. The
results of such a fit are shown in \tbn{ds-linfit}.

\bef
\centerline{\includegraphics[scale=0.8]{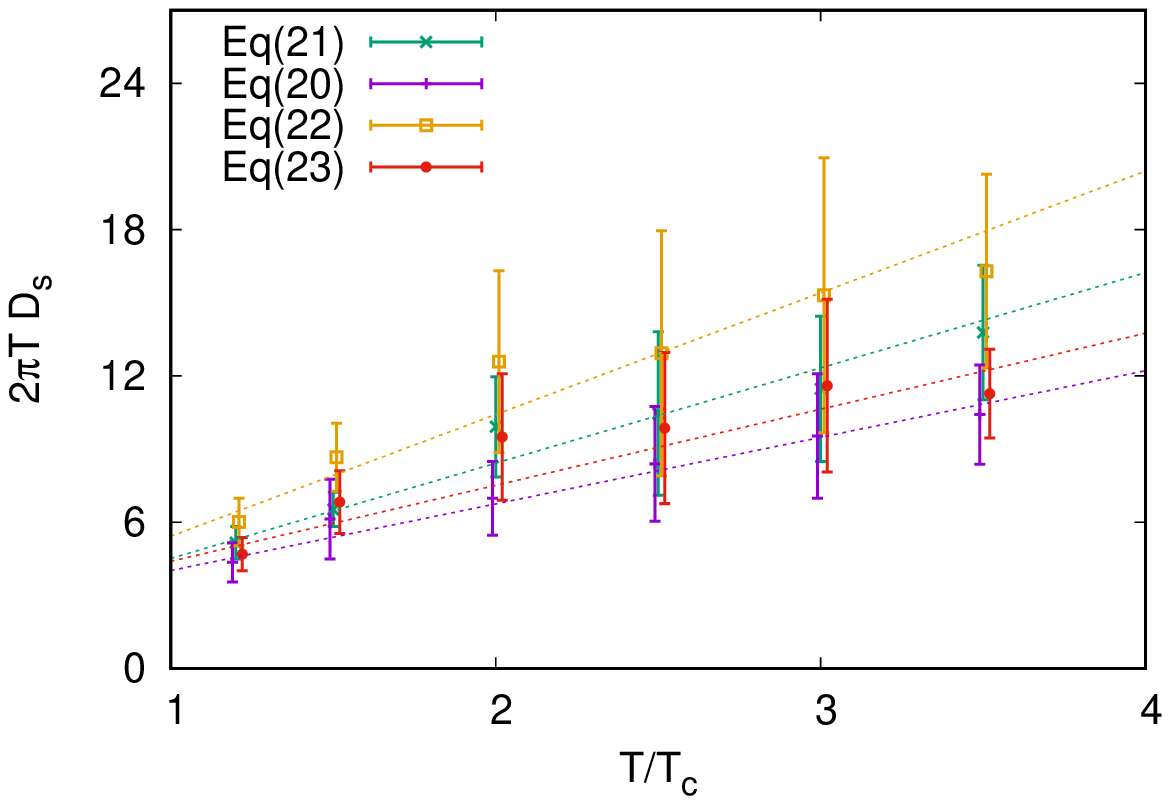}}
\caption{An estimation of the static quark diffusion
  coefficient, using \eqn{ds} and \tbn{ket}. Also shown are the best fits
  to a liear temperature dependence (\eqn{ds-linfit}).}
\eef{dsforms}

\bet
\caption{The fit parameters for \eqn{ds-linfit}}.
\begingroup
\setlength{\tabcolsep}{10pt}
\renewcommand{\arraystretch}{1.5}
\begin{center}
  \begin{tabular}{c|cccc}
    & \eqn{formb} & \eqn{formc} & \eqn{fourierb} & \eqn{fourierc} \\
    \hline
    $\alpha$ & 4.01(24) & 4.51(27) & 5.42(50) & 4.39(45) \\
    $\gamma$ & 3.91(39) & 2.73(24) & 4.99(71) & 3.12(50) \\
    \hline
\end{tabular} \end{center}
\endgroup
\eet{ds-linfit}

In \fgn{dscb} we show the estimates for $\dsc$ and $\dsb$ in the temperature
range 1.2-3 $T_c$, using \eqn{ds}, where $\kappa^c$ and $\kappa^b$ are obtained
using \eqn{om1}. The estimates for $\kb$ are taken from Ref. \cite{bb}, 
supplemented with a calculation at 3 $\tc$. 
The $BB$ correlator $\gbt$ at 3 $T_c$ at different lattice spacings,
normalized by the corresponding leading order correlator $\frac{\textstyle
  G_{\scriptscriptstyle BB}(\tau, LO)}{\textstyle g^2 C_f}$ for lattice with the
same $N_\tau$, are shown in \fgn{Bcorrfit}. We also show  
the corresponding continumm extrapolated correlator in the figure. 
The analysis for $\kb$ at 3 $T_c$ follows that used in \cite{bb}; it is
similar to the analysis for
$\ke$ outlined in \scn{tech}, except, following Ref. \cite{bb}, the scale for
the perturbative part of the $BB$ spectral function is taken to be
$\mu_{\rm fit}^{\scriptscriptstyle B} \, = \, \rm{max} \, \left[ \omega^{\frac{5}{11}} \left( \pi T \right)^{\frac{6}{11}}, \pi T \right]$ instead of \eqn{mufit}.
The results obtained for $\kb$ for the different fit forms are also shown in
\fgn{Bcorrfit}. Taking a band that includes the different fit forms, we
obtain an estimate $\kbt \, \sim$ 0.6 - 1.3 at 3 $T_c$. 

\bef
\centerline{\includegraphics[scale=0.6]{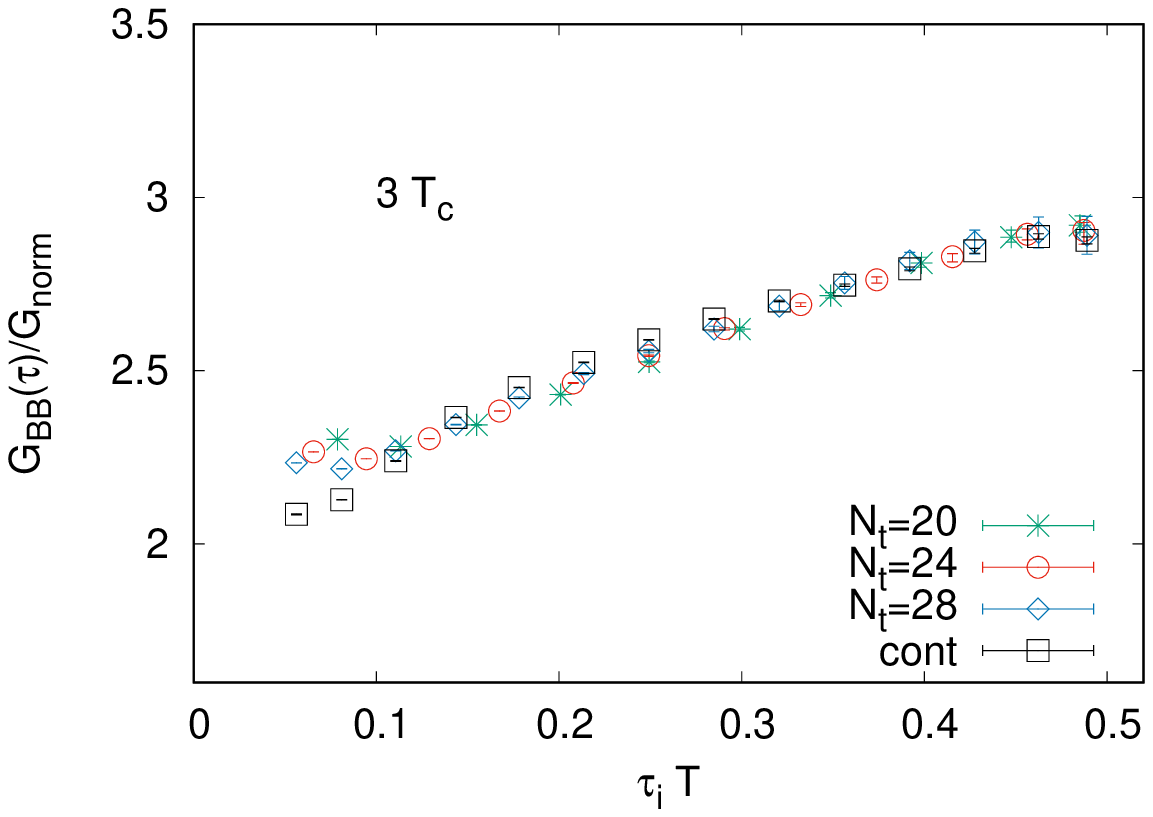}
  \includegraphics[scale=0.6]{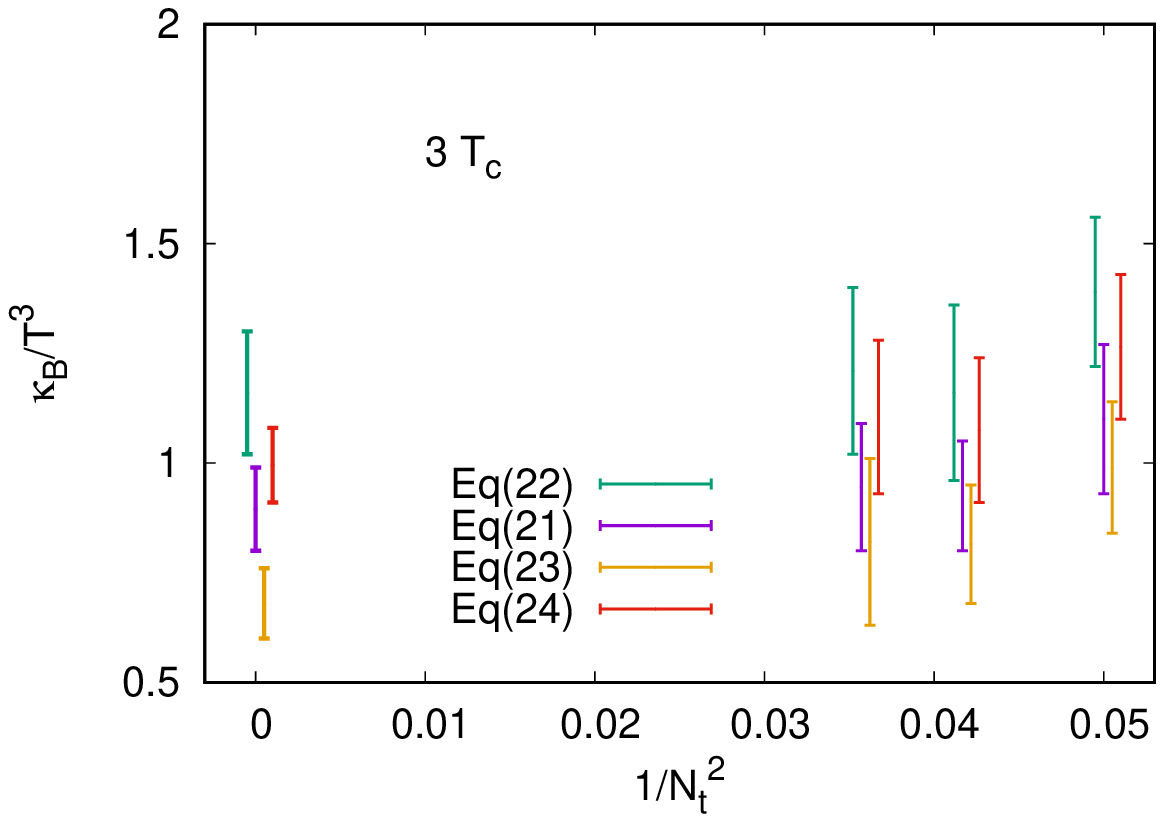}}
\caption{(Left) The correlator $\gbt$ at 3 $T_c$ calculated at different
  lattice spacings, normalized by $\gfr=\frac{\textstyle
    G_{\scriptscriptstyle BB}(\tau, LO)}{\textstyle g^2 C_f}$. Also shown is the
  continuum extrapolated correlator. (Right) Estimates of $\kbt$ at 3 $T_c$
  obtained using the different fit forms. See text for explanation.}
\eef{Bcorrfit}

Following Ref. \cite{bb}, $\langle v^2 \rangle$ was estimated from a ratio
of the susceptibilities calculated in \cite{vsq}. This gives
$\langle v^2 \rangle \approx (0.76, 0.40)$ for charm and bottom at 3 $T_c$,
respectively (the values at the lower temperatures are given in Ref.
\cite{bb}). At such temperatures, a nonrelativistic treatment of charm may be
questionable. We find that the $\om(1)$ corrections to $\kappa$, \eqn{om1},
are $\sim$ 38 \% for charm and $\sim$ 20 \% for bottom.

From the results for $\kappa^c$ and $\kappa^b$, $\dsc$ and $\dsb$ can be
obtained using \eqn{ds}. Since the range for $\kappa$ is dominated by
systematics, we simply find the range of $D_s$ by using \eqn{ds} for the
lower and upper bound of the range for $\kappa$. The results for $\dscb$
obtained this way are given in \tbn{dscb} and shown in \fgn{dscb}.

\bet
\caption{The estimates for $\dscb$, from \eqn{ds} and \eqn{om1}.}
\begingroup
\setlength{\tabcolsep}{10pt}
\renewcommand{\arraystretch}{1.5}
\begin{center}
  \begin{tabular}{c|cccc}
    & 1.2 $T_c$ & 1.5 $T_c$ & 2 $T_c$ & 3 $T_c$ \\
    \hline
    $2 \pi \, T \, \dsc$ & 2.9 - 5.1 & 3.5 - 6.6 & 4.0 - 9.9 & 5.1 - 11.4 \\
    $2 \pi \, T \, \dsb$ & 3.3 - 5.6 & 4.0 - 7.6 & 4.7 - 11.2 & 5.9 - 13.1 \\
    \hline
\end{tabular} \end{center}
\endgroup
\eet{dscb}


\begin{thebibliography}{99}
\bibitem{lonlo}
  Note that the EE correlator at LO does not have a diffusion term; the 
  leading order estimate for $\ket$ comes from the EE correlator 
  calculated at NLO.
\bibitem{caution}
  Here we have done a simple $\chi^2$ analysis, treating the error band
  in \tbn{ket} as a statistical 1$\sigma$ band. The error band in the table is
  dominated by systematic errors. So while the fit does 
  indicate that the data is inconsistent with a flat behavior with temperature,
  one should not attribute the standard 1$\sigma$ interpretation to the
  error bars quoted for the fit parameters.
\end{thebibliography}
\end{document}